\documentstyle[prb,aps,epsfig,floats]{revtex}

\begin{document}

\draft
\twocolumn[\hsize\textwidth\columnwidth\hsize\csname
@twocolumnfalse\endcsname

\title{Density-functional study of the structure and stability
       of ZnO surfaces}

\author{B. Meyer and Dominik Marx}

\address{Lehrstuhl f\"ur Theoretische Chemie,\\
         Ruhr-Universit\"at Bochum, 44780 Bochum, Germany}
\date{\today}
\maketitle

\begin{abstract}

An extensive theoretical investigation of the nonpolar (10$\bar{1}$0) and
(11$\bar{2}$0) surfaces as well as the polar zinc terminated (0001)--Zn and
oxygen terminated (000$\bar{1}$)--O surfaces of ZnO is presented. Particular
attention is given to the convergence properties of various parameters such as
basis set, k--point mesh,  slab thickness, or relaxation constraints within
LDA and PBE pseudopotential calculations using both plane wave and mixed basis
sets. The pros and cons of different approaches to deal with the stability
problem of the polar surfaces are discussed. Reliable results for the
structural relaxations and the energetics of these surfaces are presented and
compared to previous theoretical and experimental data, which are also
concisely reviewed and commented.

\end{abstract}


\vskip2pc]
\narrowtext


\section{Introduction}
\label{sec:intro}

The II--VI semiconductor ZnO has become a frequently studied material in
surface science because of its wide range of technological applications.
ZnO is a basic material for varistors, thyristors and optical coatings.
In addition, its direct band gap makes it an interesting candidate for blue
and UV emitting LEDs and laser diodes.\cite{laser}  The electronic and
structural properties of the ZnO surfaces are in particular important in its
applications as chemical sensor in gas detecting systems and as catalyst for
hydrogenation and dehydrogenation reactions. In combination with Cu particles
at the surface, ZnO is a very efficient catalyst for the methanol
synthesis\cite{catalysis} where it is employed in industrial scale. The
mechanism behind the enhanced catalytic activity when combined with Cu is
poorly understood. However, before this interesting interplay between the
ZnO substrate and the Cu particles can be addressed, a thorough understanding
of the underlying clean ZnO surfaces is necessary.

\begin{figure}[t]
\noindent
\epsfxsize=246pt
\centerline{\epsffile{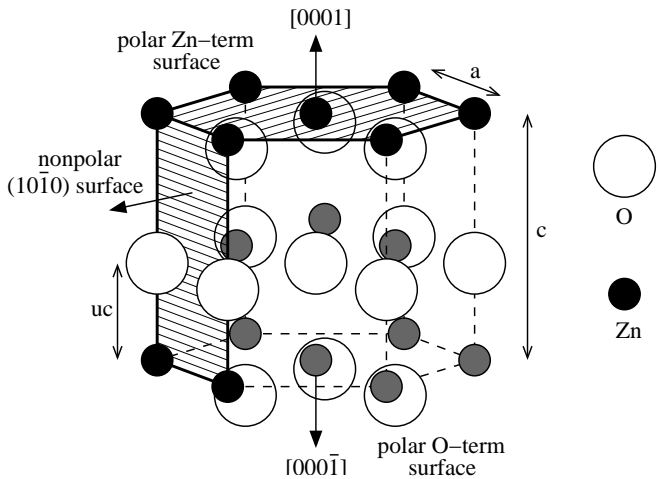}}
\vspace{4pt}
\caption{\label{fig:bulk}
Wurtzite structure (B4) of ZnO with the polar zinc terminated (0001)--Zn, the
polar oxygen terminated (000$\bar{1}$)--O, and the nonpolar (10$\bar{1}$0)
surfaces.}
\end{figure}

From a physical/chemical point of view, ZnO is a very interesting material
because of the mixed covalent/ionic aspects in the chemical bonding. ZnO
crystallizes in the hexagonal wurtzite structure (B4) which consists of
hexagonal Zn and O planes stacked alternately along the $c$--axis (see
Fig.~\ref{fig:bulk}). Anions and cations are 4--fold coordinated,
respectively, like in the closely related zincblende structure. A tetrahedral
coordinated bulk structure is typical for rather covalent semiconductors. On
the other hand, ZnO shows great similarities with ionic insulators such as
MgO.\cite{cox}  This is why ZnO is often called the `ionic extreme' of
tetrahedral coordinated semiconductors.

Wurtzite crystals are dominated by four low Miller index surfaces: the
nonpolar (10$\bar{1}$0) and (11$\bar{2}$0) surfaces and the polar zinc
terminated (0001)--Zn and the oxygen terminated (000$\bar{1}$)--O surfaces
(see Fig.~\ref{fig:bulk}). By ion sputtering and annealing at not too high
temperatures all four surfaces can be prepared in a bulk terminated,
unreconstructed state, where the surface atoms only undergo symmetry
conserving relaxations. A typical p(1$\times$1) pattern is observed in
low-energy electron diffraction (LEED) and other diffraction
experiments.\cite{duke1,duke2,duke3,noguera}  Although in a recent
He--scattering experiment\cite{woell} it was shown that O--terminated
(000$\bar{1}$) surfaces with p(1$\times$1) LEED patterns are usually hydrogen
covered whereas the clean O--terminated surface exhibits a (3$\times$1)
reconstruction, we will focus in this study on the clean, unreconstructed
surfaces of ZnO.

In the present paper, we investigate all four main crystal terminations of
ZnO. The fully relaxed geometric structures and the surface/cleavage energies
have been calculated using a first-principles density-functional (DFT)
method. We have employed both, a local-density (LDA) and a
generalized-gradient approximation (GGA) functional. We will discuss the
relative stability of the four surfaces and how the surface relaxations of the
nonpolar faces are connected to the covalency/ionicity of the chemical bond in
ZnO. Finally, a detailed comparison with existing theoretical and
experimental results will be given.

The nonpolar (10$\bar{1}$0) surface of ZnO has been the focus of several
experimental and theoretical studies. However, the form of the relaxation
of the surface atoms is still very controversial. Duke et al.\cite{duke2}
concluded from their best LEED analysis\cite{comment1} that the top-layer
zinc ion is displaced downwards by $\Delta d_\perp$(Zn)=$-$0.45$\pm$0.1\,{\AA}
and likewise the top-layer oxygen by
$\Delta d_\perp$(O)=$-$0.05$\pm$0.1\,{\AA}, leading to a tilt of the Zn--O
dimer of 12$^\circ\pm$5$^\circ$. No compelling evidence for lateral
distortions within the first layer or for second-layer relaxations were
obtained, but small improvements could be achieved by assuming a lateral
displacement of the Zn ion toward oxygen by 
$\Delta d_\parallel$(Zn)=0.1$\pm$0.2\,{\AA}.\cite{comment2}
The strong inward relaxation of the Zn ion was later confirmed by G\"opel
at al.\cite{goepel} in an angle-resolved photoemission experiment. By
comparing the relative position of a particular surface state with its
theoretically predicted geometry dependence, a Zn displacement downwards by
$\Delta d_\perp$(Zn)=$-$0.4\,{\AA} was concluded.

In contrast, Jedrecy et al.\cite{jedrecy1} found best agreement with their
gracing incidence X--ray diffraction data (GIXD) for a structural model where
the top-layer zinc atom is displaced downwards by only
$\Delta d_\perp$(Zn)=$-$0.06$\pm$0.02\,{\AA} and shifted toward oxygen by
$\Delta d_\parallel$(Zn)=0.05$\pm$0.02\,{\AA}. However, for their samples they
observed a high density of steps and from their best-fit model they predict
rather high vacancy concentrations in the first two surface layers
with occupancy factors of 0.77$\pm$0.02 and 0.90$\pm$0.04 for the first and
second layer, respectively. On the other hand, Parker et al.\cite{parker}
reported scanning tunneling microscopy (STM) images of the nonpolar
(10$\bar{1}$0) surfaces with atomic resolution where large flat terraces are
found and no defects are visible in areas as large as 11$\times$14 surface
unit cells. Due to the small scattering contribution, the position of oxygen
could not be determined very accurately in the GIXD experiment of
Ref.~\onlinecite{jedrecy1}. The result of the best fit was that O relaxes
further toward the bulk than Zn with
$\Delta d_\perp$(O)=$-$0.12$\pm$0.06\,{\AA}. This would be very unusual
since to our knowledge no (10$\bar{1}$0) wurtzite or (110) zincblende surface
structure has been reported where the surface dimers tilts with the cation
above the anion.

First theoretical investigations of the (10$\bar{1}$0) surface were done using
empirical tight-binding (TB) models. With two very different TB models Wang
and Duke\cite{wang} found a strong displacement of
$\Delta d_\perp$(Zn)=$-$0.57\,{\AA}, whereas Ivanov and Pollmann\cite{ivanov}
obtained an almost bulk-like surface geometry. A recent calculation with
atomistic potentials based on a shell model\cite{catlow} predicted
$\Delta d_\perp$(Zn)=$-$0.25\,{\AA} and a rather strong upward relaxation of
the second-layer Zn of +0.165\,{\AA}.

Several ab-initio studies (DFT-LDA,\cite{schroer1}  Hartree-Fock
(HF),\cite{jaffe1} and a hybrid HF and DFT method using the B3LYP
functional\cite{wander1}) employing Gaussian orbitals as basis functions to
solve the electronic structure problem favor small inward relaxations of Zn
and small tilts of the ZnO--dimers of 2$^\circ$--5$^\circ$. However,
it is questionable, if these studies represent fully converged results.
There is only one recent first-principles DFT-LDA calculation using plane
waves\cite{filippetti} where larger relaxations with a tilt of 11.7$^\circ$
were obtained.

The nonpolar (11$\bar{2}$0) ZnO surface has been less frequently studied
than its (10$\bar{1}$0) counterpart. The two tight-binding
models\cite{wang,ivanov} predicted the same relaxation behavior for the
(11$\bar{2}$0) as for the (10$\bar{1}$0) surface: Wang and Duke\cite{wang}
found a strong zinc displacement of $\Delta d_\perp$(Zn)=$-$0.54\,{\AA}
toward the bulk whereas the TB model of Ivanov and Pollmann preferred an
almost bulk-like surface structure. With a first-principles hybrid
B3LYP method Wander and Harrison\cite{wander2} found much smaller
relaxations for the (11$\bar{2}$0) surface than for the (10$\bar{1}$0)
face, but not all degrees of freedom were relaxed in this study. To our
knowledge there has been no quantitative experimental investigation.

\begin{table}[!b]
\noindent
\begin{center}
\begin{minipage}[c]{246pt}
\def\arraystretch{1.2}
\begin{tabular}{lccc}
                  &   LDA               &   PBE               &  Expt. \\
\hline
$a$ [\AA]         &  3.193 ($-$1.7\,\%) &  3.282 ($+$1.0\,\%) &  3.250 \\
$c$ [\AA]         &  5.163 ($-$0.8\,\%) &  5.291 ($+$1.6\,\%) &  5.207 \\
$c$/$a$           &  1.617              &  1.612              &  1.602 \\
$u$               &  0.3783             &  0.3792             &  0.3825 \\
$B_0$ [GPa]       &  161                &  128                &  143 \\
$p_{\rm T}$ [GPa] &  9.0                &  11.8               &  9.0--9.5
\end{tabular}
\end{minipage}
\end{center}
\caption{\label{tab:bulk}
Computed and experimental values of the structural parameters for bulk
ZnO. $a$ and $c$ are the lattice constants, $u$ is an internal coordinate of
the wurtzite structure which determines the relative position of the anion
and cation sublattice along the $c$ axis, $B_0$ is the bulk modulus, and
$p_{\rm T}$ is the transition pressure between the wurtzite (B4) and
rocksalt (B1) structure of ZnO. Experimental values are from
Refs.~\protect\onlinecite{exp-bulk1,exp-bulk2,exp-bulk3}. Relative deviations
from experiment are given in parenthesis.}
\end{table}

\begin{figure}[!t]
\noindent
\epsfxsize=246pt
\centerline{\epsffile{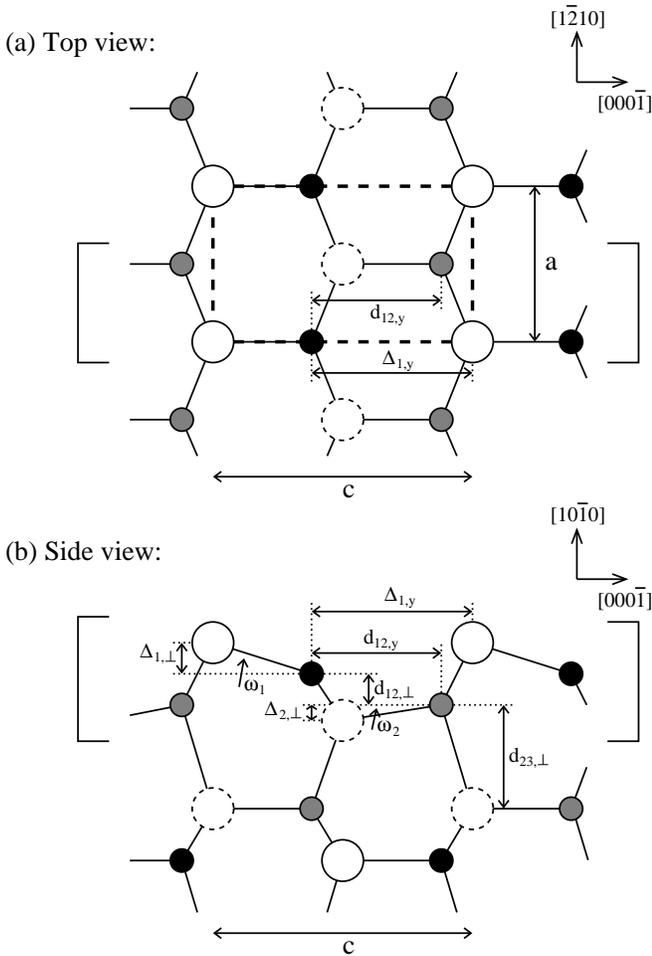}}
\vspace{4pt}
\caption{\label{fig:1010}
Schematic diagram of the surface geometry and the independent structural
parameters of the nonpolar (10$\bar{1}$0) surface. The brackets indicate the
two atomic layers shown in top and side view. Open and filled symbols are the
O and Zn ions, respectively, and the solid lines represent nearest-neighbor
bonds. The atoms in the first layer are shown by solid/black, second layer
atoms by dashed/shaded circles. The surface unit cell is indicated by dashed
lines.}
\end{figure}

Coming to the polar surfaces, we encounter the fundamental problem that in
an ionic model these surfaces are unstable and should not exist. They are
so-called `Tusker type 3' surfaces,\cite{tusker} and with simple
electrostatic arguments it can be shown that the surface energy diverges for
such a configuration.\cite{tusker}  To stabilize the polar surfaces a
rearrangement of charges between the O-- and the Zn--terminated surfaces needs
to take place in which the Zn--terminated side becomes less positively charged
and the O--terminated face less negative. In fact, most polar surfaces show
massive surface reconstructions or exhibit facetting to accommodate the charge
transfer.\cite{noguera}  Also randomly distributed vacancies, impurity atoms
in the surface layers, or the presence of charged adsorbates are possible
mechanisms to stabilize polar surfaces. However, the polar ZnO surfaces are
remarkably stable, and many experiments suggest that they are in an
unreconstructed, clean and fully ordered state.\cite{noguera}  Despite
many investigations it is still an open question how the polar ZnO surfaces
are stabilized.\cite{noguera}

Assuming clean and unreconstructed surfaces, the reduction in surface charge
density can only occur from a redistribution of the electrons. Negative
charge has to be transferred from the O--terminated face to the Zn--terminated
side, leading to partially occupied bands at the surface. This so called
`metallization of the surface' has been used by all previous ab-initio
calculations~\cite{wander3,carlsson,noguera} to model the polar ZnO surfaces
and will also be employed in the present study. However, whether or not the
surfaces are metallic will depend on the width of the partially occupied
bands. From another point of view, if the polar surfaces were stabilized by
vacancies, defects or adsorbates, many defect states would be created. Now
if we think of somehow averaging over the surface, the defect states would
form a partially filled band. In this sense, the `metallization' may be
regarded as some `mean-field' description for a situation where many defect
states are present.

Several attempts have been made to determine the layer relaxations of the
unreconstructed polar surfaces. In an early dynamical LEED analysis Duke
and Lubinsky\cite{duke4} found an outer Zn--O double-layer spacing of
$d_{12}$=0.607\,{\AA} for the Zn--terminated surface and $d_{12}$=0.807\,{\AA}
for the O--terminated face. Unfortunately, this analysis was based on an
early bulk structure of ZnO, see Ref.~\onlinecite{wyckoff}, in which the bulk
double-layer spacing was assumed to be 0.807\,{\AA} instead of 0.612\,{\AA}.

For the Zn--terminated surface, it was concluded from the comparison of
X--ray photo-diffraction (XPD) data with scattering simulations\cite{xpd1}
that any inward relaxation of the surface Zn layer can be ruled out.
Coaxial impact-collision ion-scattering spectroscopy\cite{caiciss} (CAICISS)
proposed an expansion of $d_{12}$ by +0.35\,{\AA}. Also an expansion of
$d_{12}$ by +0.05\,{\AA} for the Zn--terminated surface was found in an GIXD
measurement.\cite{jedrecy2} In this experiment, the X--ray data could best be
fitted by assuming a random removal of 1/4 of the Zn atoms in the surface
layer. On the other hand, from the shadowing and blocking edges of a
low-energy alkali ion scattering\cite{overbury} (LEIS) experiment no evidence
for substantial quantities of point defects in the Zn--terminated as well as
in the O--terminated surface was found.

For the O--terminated surface, it was concluded from LEIS\cite{overbury} that
the Zn--O double-layer spacing $d_{12}$ is close its bulk value. An XPD
study\cite{xpd2} found a contraction of 25\,\% of $d_{12}$, but like in the
LEED analysis\cite{duke4}, the wrong bulk structure of
Ref.~\onlinecite{wyckoff} was used in the scattering simulations.
A GIXD measurement\cite{jedrecy2} predicted also an inward relaxation
of the topmost O--layer by $-$0.33\,{\AA} and an outward relaxation of the
underlying Zn--plane by +0.08\,{\AA}. The occupancy probabilities were fitted,
resulting in 1.3(!)\ and 0.7 for the first bilayer O and Zn, respectively.
After considerably improved sample preparation was achieved, the same authors
reinvestigated the O--terminated polar surface.\cite{jedrecy3}
Best agreement with their GIXD data was now found for a structural model where
both, the upper O and Zn planes relax inwards by $-$0.19$\pm$0.02\,{\AA} and
$-$0.07$\pm$0.01\,{\AA}, respectively, with occupancy factors of 1.0 in the
oxygen plane and 0.75$\pm$0.03 in the underlying Zn plane. The inward
relaxation of the O--layer has been confirmed by another surface X--ray
diffraction measurement\cite{wander3} where $\Delta
d_{12}$=$-$0.24$\pm$0.06\,{\AA} and $\Delta d_{23}$=+0.04$\pm$0.05\,{\AA} was
obtained.

Ab-initio calculations on polar slabs\cite{wander3,carlsson,noguera} predict
consistently for both surface terminations contractions for the first Zn--O
double-layer distance, with a larger inward relaxation at the O--terminated
surface.

In view of the above discussed discrepancies between different experimental
and theoretical investigations, it is our aim to provide a consistent set of
fully converged calculations for the four main ZnO surfaces. We attempt to
overcome the restrictions of previous theoretical studies such that the
current study can be regarded as a reference for perfectly ordered,
defect-free surfaces. An accurate set of uniform theoretical data may then
allow to discuss the differences between theory and experiment in terms of
deviations between the model of ideal, unreconstructed surfaces as assumed in
the ab-initio simulations and the structure of the surfaces occurring in
nature. In particular, for the polar surfaces this may give new insight into
how these surfaces are stabilized.


\section{Theoretical details}
\label{sec:theorie}

\subsection{Method of calculation and bulk properties}
\label{sec:method}

We have carried out self-consistent total-energy calculations within the
framework of density-functional theory (DFT).\cite{hks}  The exchange and
correlation effects were treated within both, the local-density approximation
(LDA),\cite{ca,sic}  and the generalized-gradient approximation (GGA) where
we used the functional of Perdew, Becke, and Ernzerhof\cite{pbe} (PBE).

Two different pseudopotential schemes were applied: For the study of the
nonpolar surfaces we used pseudopotentials of the Vanderbilt ultrasoft
type.\cite{van-usp}  The electronic wave functions were expanded in a plane
wave basis set including plane waves up to a cut-off energy of 25\,Ry.
A conjugate gradient technique as described in Ref.~\onlinecite{ksv} was
employed to minimize the Kohn-Sham total energy functional.

For the calculations on the polar surfaces we used normconserving
pseudopotentials\cite{van-pp} together with a mixed-basis consisting of
plane waves and non-overlapping localized orbitals for the O--$2p$ and
the Zn--$3d$ electrons.\cite{mb}  A plane-wave cut-off energy of 20\,Ry
was sufficient to get well converged results. To improve convergence in the
presence of partly occupied bands, a Gaussian broadening\cite{kmh} with a
smearing parameter of 0.1\,eV was included. For several configurations
representing nonpolar surfaces we repeated the calculations with the
mixed-basis approach. No significant differences compared to the results 
from the ultrasoft-pseudopotential method could be seen.

\begin{figure}[!t]
\noindent
\epsfxsize=246pt
\centerline{\epsffile{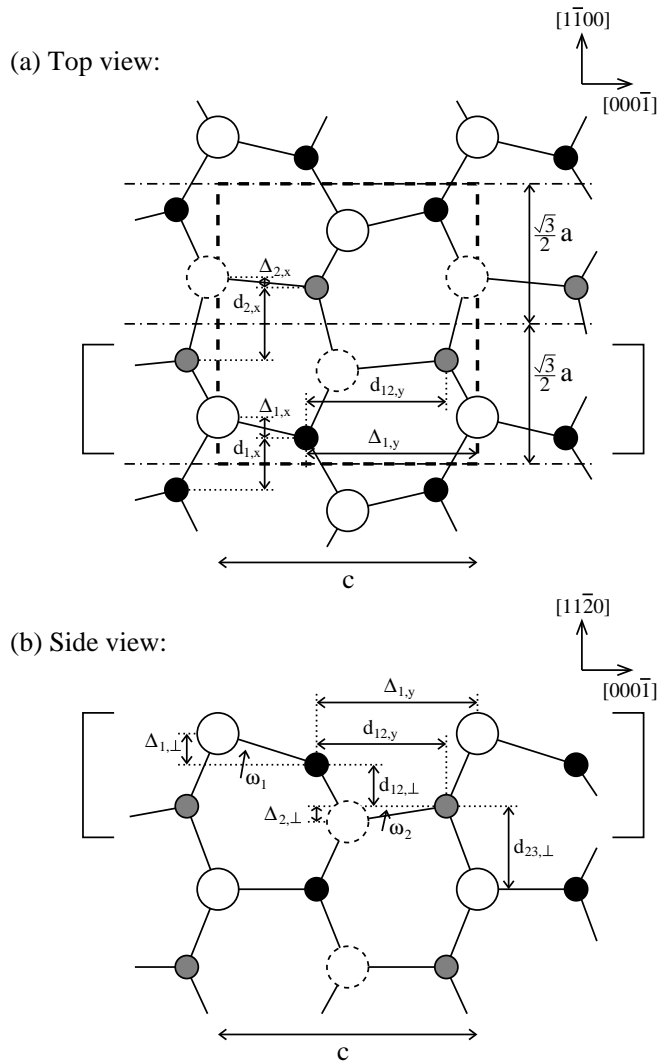}}
\vspace{4pt}
\caption{\label{fig:1120}
Schematic top and side view of the surface geometry of the nonpolar
(11$\bar{2}$0) surface. The same representation as in
Fig.~\protect\ref{fig:1010} is used. The glide planes are indicated by
dashed-dotted lines.}
\end{figure}

\begin{figure}[!t]
\noindent
\epsfxsize=246pt
\centerline{\epsffile{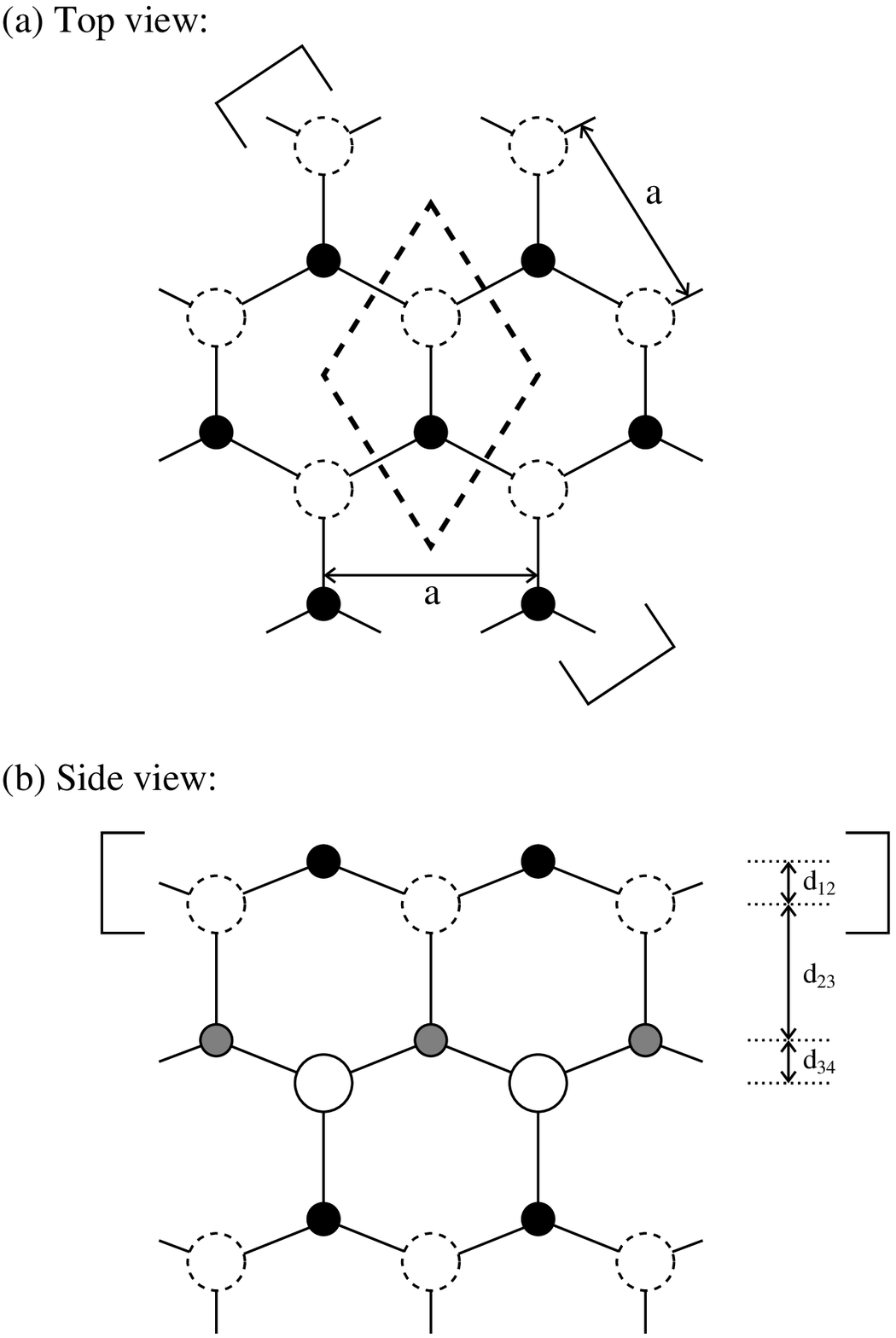}}
\vspace{4pt}
\caption{\label{fig:0001}
Schematic top and side view of the polar Zn--terminated (0001) surface.
The same representation as in Fig.~\protect\ref{fig:1010} is used.}
\end{figure}

It is a well known shortcoming of LDA and GGA that both predict the
Zn--$d$ bands to be roughly 3\,eV too high in energy as compared to
experiment.\cite{goepel,girard}  In consequence, the Zn--$d$ states hybridize
stronger with the O--$p$ valence bands, thereby shifting them unphysically
close to the conduction band. The underestimate for the band gap is therefore
even more severe in ZnO than in other semiconductors. In our calculations
we obtained band gaps of 0.78\,eV and 0.74\,eV with LDA and PBE, respectively,
as opposed to the experimental value of 3.4\,eV. The band gap and the position
of the Zn--$d$ bands can be improved significantly, if a self-interaction
correction (SIC) is used.\cite{sic}  Usually SIC calculations are very
demanding, but if the SIC effects are incorporated into the
pseudopotential\cite{sic-pp}, the additional calculational cost is modest.
Unfortunately, the SIC pseudopotential scheme does not improve the structural
properties of ZnO\cite{sic-pp} and also causes some problems when accurate
atomic forces are needed.\cite{diss-vogel}  
Therefore, and since we are mostly interested in accurate relaxed geometries
of the surfaces and not so much in their electronic structure, we have
omitted the use of SIC in our calculations.

The computed structural parameters for bulk ZnO are shown in
Table~\ref{tab:bulk}. Mixed-basis and ultrasoft-pseudopotential calculations
give the same results within the accuracy displayed in Table~\ref{tab:bulk}.
As is typical for the functionals, LDA underestimates the lattice constants
by 1--2\,\%, and GGA overestimates them by roughly the same amount. The
$c/a$--ratio strongly influences the internal parameter $u$.
If $u\!=\!1/4+a^2/3c^2$, all nearest-neighbor bonds are equal.
Since the $c/a$--ratio is slightly overestimated in our calculations, we
get $u$--values that are slightly smaller than observed in experiment.

The construction of appropriate supercells for the study of the surfaces
will be detailed in the following subsection. All atomic configurations
were fully relaxed by minimizing the atomic forces using a variable-metric
scheme.\cite{numrec} Convergence was assumed when the forces on the ions
were less than 0.005\,eV/\AA.

\subsection{Surfaces, slab structures, and the stability problem}
\label{sec:surfgeom}

All surfaces were represented by periodically repeated slabs consisting of
several atomic layers and separated by a vacuum region of 9.4 to 12.4\,\AA.
For the polar surfaces a dipole correction\cite{bengtsson,bm} was used to
prevent artificial electrostatic interactions between the repeated
units. To simulate the underlying bulk structure, the slab lattice constant
in the direction parallel to the surface was always set equal to the
theoretical equilibrium bulk value (see Table~\ref{tab:bulk}).

The nonpolar surfaces are obtained by cutting the crystal perpendicular to the
hexagonal Zn-- and O--layers (see Fig.~\ref{fig:bulk}). In both cases, for the
(10$\bar{1}$0) and the (11$\bar{2}$0) planes, two equivalent surfaces are
created so that always stoichiometric slabs with the same surface termination
on top and on bottom can be formed.

The (10$\bar{1}$0) surface geometry is sketched in Fig.~\ref{fig:1010}. Each
surface layer contains one ZnO dimer. The dimers form characteristic rows
along the [1$\bar{2}$10] direction which are separated by trenches. Slabs with
4--20 atomic layers were used, thus containing up to 40 atoms, and the
Brillouin-zone of the supercell was sampled with a (4$\times$2$\times$2)
Monkhorst-Pack\cite{mp} k--point grid. No differences were found when going
to a (6$\times$4$\times$2) mesh.

The surface layers of the (11$\bar{2}$0) surface are built up by two ZnO
dimers which form zig-zag lines along the surface (see Fig.~\ref{fig:1120}).
The two dimers are equivalent and are related by a glide plane symmetry. This
symmetry is not destroyed by the atomic relaxations of the
surface.\cite{duke3}  The slabs in our calculations were built of 4--8
atomic layers with up to 32 atoms, and a (2$\times$2$\times$2)
Monkhorst-Pack\cite{mp} k--point mesh was used. Again, a denser
(4$\times$4$\times$2) mesh did not alter the results.

\begin{figure}[!b]
\noindent
\epsfxsize=246pt
\centerline{\epsffile{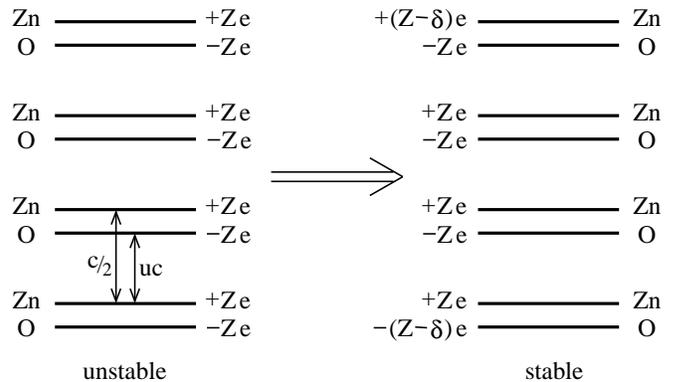}}
\vspace{4pt}
\caption{\label{fig:slab}
Schematic illustration of the stacking sequence of the polar slabs. A charge
transfer of $\delta = (1\!-\!2u)\,c\approx Z/4$ has to occur to stabilize
the polar surfaces.}
\end{figure}

\begin{figure}[!t]
\noindent
\epsfxsize=246pt
\centerline{\epsffile{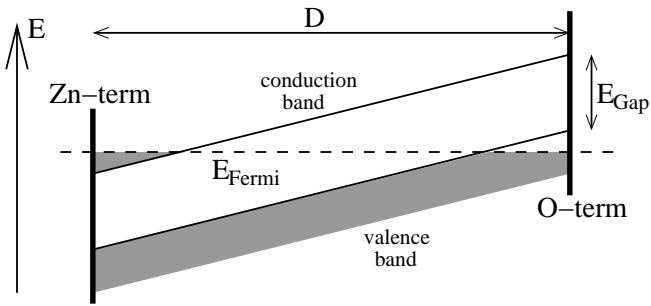}}
\vspace{4pt}
\caption{\label{fig:band}
Schematic illustration of the band structure after electrons have moved from
the O-- to the Zn--terminated surface of the slab. Depending on the band gap
and the thickness $D$ of the slab, a residual electric field remains inside
the slab.}
\end{figure}

Cleaving the crystal perpendicular to the $c$--axis (see Fig.~\ref{fig:bulk})
always creates simultaneously a Zn-- and an O--terminated polar (0001) and
(000$\bar{1}$) surface, respectively. If we only consider cuts where the
surface atoms stay 3--fold coordinated, all slabs representing polar surfaces
are automatically stoichiometric and are inevitably Zn--terminated on one side
and O--terminated on the other side. Figure~\ref{fig:0001} sketches the
characteristic sequence of Zn--O double-layers of the polar slabs. In our
calculations slabs with 4--20 Zn--O double-layers were used, thus containing
8--40 atoms. k--point convergence was achieved with a (6$\times$6$\times$1)
Monkhorst-Pack\cite{mp} grid, and tests with up to (12$\times$12$\times$1)
k--points were made.

Each Zn--O double-layer in Fig.~\ref{fig:0001} exhibits a dipole moment
perpendicular to the surface. If we assume for simplicity a purely ionic model 
for ZnO and assign the fixed formal charges $+Ze$ and $-Ze$ to the Zn-- and
O--ions, respectively, then a slab of $N$ double-layers will exhibit a dipole
moment of $m\!=\!N\,Ze\,(1\!-\!2u)\,c/2$ (see Fig.~\ref{fig:slab}). This
corresponds to a spontaneous polarization of $P_{\rm s}\!=\!Ze\,(1\!-\!2u)$
which is independent of the thickness of the slab. If the external electric
field is zero, inside the slab an electric field of $E\!=\!-4\pi P_{\rm s}$
will be present. Therefore, no matter how thick we choose our slab, the inner
part will {\em never} become bulk-like, and the surface energy, defined as
the difference between the energy of the slab and the energy of the same
number of atoms in the bulk environment, will {\em diverge} with slab
thickness.\cite{tusker}  Thus, the polar surfaces are {\em not stable}.

On the other hand, it can easily be seen that if we modify the charge in the
top and bottom layer of the slab from $\pm Ze$ to $\pm (Z\!-\!\delta)e$ with
$\delta\!=\!(1\!-\!2u)\,Z \approx Z/4$, then the dipole moment of the slab
will become {\em independent} of the slab thickness and the internal electric
field {\em vanishes}. This charge transfer is equivalent to applying an
external dipole which compensates the internal electric field.

For most polar surfaces the rearrangement of the charges is accomplished by a
modification of the surface layer composition with respect to the bulk. If
this does not occur, the internal electric field will `tilt' the band
structure by which the upper edge of the valence band close to the
O--terminated surface will become higher in energy than the lower edge of the
conduction band at the Zn--terminated face (see Fig.~\ref{fig:band}). The slab
can now lower its energy (thereby reducing the internal electric field) by
transferring electrons from the valence band at the O--terminated side to the
conduction band at the Zn--terminated face. This will happen `automatically'
in any self-consistent electronic structure calculation that makes use of a
slab geometry. This is what is usually referred to as `the metallization of
polar surfaces'.

However, one problem still remains: electrons move from the O-- to the
Zn--terminated surface until the upper valence band edge at the O--terminated
side has reached the same energy as the lower edge of the conduction band at
the Zn--terminated face as sketched in Fig.~\ref{fig:band}. In this situation,
the internal electric field is not {\em fully} removed for a finite slab with
thickness $D$. The residual electric field depends on the band gap and
vanishes only with 1/$D$. In our calculations we found that for slabs with up
to 6 Zn--O double-layers the residual electric field is still so strong that
the slabs are not stable. There is no energy barrier when the O-- and
Zn--layers are shifted simultaneously and rigidly toward each
other. Therefore, to get good converged results for the surface geometries and
energies very thick slabs have to be used which makes the investigation of the
polar surfaces computationally very demanding. Ideally, one should calculate
all quantities of interest for different slab thicknesses $D$ and extrapolate
the results to 1/$D \longrightarrow 0$. In the present study we obtained the
relaxations of the surface layers (see Fig.~\ref{fig:relax}) as well as the
cleavage energy of the polar surfaces (see Fig.~\ref{fig:energy}) by
extrapolating the results of slab calculations containing up to 20 Zn--O
double-layers.

\begin{table}
\noindent
\begin{center}
\begin{minipage}[c]{246pt}
\def\arraystretch{1.2}
\begin{tabular}{ccccc}
   & \multicolumn{2}{c}{(10$\bar{1}$0) surface}
   & \multicolumn{2}{c}{(11$\bar{2}$0) surface} \\
   & LDA & PBE & LDA & PBE \\ \hline
$\Delta_{1,\perp}$
   & $+$0.106\,$a$ & $+$0.100\,$a$ & $+$0.076\,$a$ & $+$0.073\,$a$ \\
$\Delta_{2,\perp}$
   & $-$0.041\,$a$ & $-$0.038\,$a$ & $-$0.016\,$a$ & $-$0.015\,$a$ \\
Bulk
   & \multicolumn{2}{c}{0.0} & \multicolumn{2}{c}{0.0} \\[4pt]
$\Delta_{1,y}$
   & 0.6531\,$c$ & 0.6539\,$c$ & 0.6506\,$c$ & 0.6516\,$c$ \\
$\Delta_{2,y}$
   & 0.6243\,$c$ & 0.6231\,$c$ & 0.6230\,$c$ & 0.6221\,$c$ \\
Bulk
   & \multicolumn{2}{c}{$(1\!-\!u)\,c$} &
     \multicolumn{2}{c}{$(1\!-\!u)\,c$} \\[4pt]
$\Delta_{1,x}$
   &        &        &  0.083\,$a$ &  0.077\,$a$ \\
$\Delta_{2,x}$
   &        &        &  0.020\,$a$ &  0.016\,$a$ \\
Bulk
   &        &        & \multicolumn{2}{c}{0.0} \\[4pt]
$d_{12,\perp}$
   & 0.1445\,$a$ & 0.1447\,$a$ & 0.4093\,$a$ & 0.4089\,$a$ \\
$d_{23,\perp}$
   & 0.6328\,$a$ & 0.6337\,$a$ & 0.5215\,$a$ & 0.5222\,$a$ \\
Bulk
   & \multicolumn{2}{c}{$\frac{\sqrt{3}}{6}a$ / $\frac{\sqrt{3}}{3}a$}
   & \multicolumn{2}{c}{$a/2$} \\[4pt]
$d_{12,y}$
   & 0.5355\,$c$ & 0.5357\,$c$ & 0.5259\,$c$ & 0.5266\,$c$ \\
$d_{23,y}$
   & 0.5013\,$c$ & 0.5017\,$c$ & 0.5014\,$c$ & 0.5009\,$c$ \\
Bulk
   & \multicolumn{2}{c}{$c/2$} & \multicolumn{2}{c}{$c/2$} \\[4pt]
$d_{1,x}$
   &        &        & 0.4381\,$a$ & 0.4399\,$a$ \\
$d_{2,x}$
   &        &        & 0.5515\,$a$ & 0.5556\,$a$ \\
Bulk
   &        &        & \multicolumn{2}{c}{$\frac{\sqrt{3}}{3}a$}
\end{tabular}
\end{minipage}
\end{center}
\caption{\label{tab:nonpolar}
Summary of the structural relaxations of the first two surface layers for
the nonpolar (10$\bar{1}$0) and (11$\bar{2}$0) surfaces. The definitions
of the independent structural parameters are shown in
Fig.~\protect\ref{fig:1010} and \ref{fig:1120}. All relaxations are given
in fractions of the theoretical bulk lattice constants $a$ and $c$ (see
Table~\protect\ref{tab:bulk}). The rows labeled 'Bulk' are the corresponding
values for the unrelaxed surface.}
\end{table}

\begin{table}
\noindent
\begin{center}
\begin{minipage}[c]{246pt}
\def\arraystretch{1.2}
\begin{tabular}{lcccc}
 & $\omega$ & $C_{\rm B,\parallel}$ & $C_{\rm B}$(Zn) & $C_{\rm B}$(O)\\
\hline
LDA, this study     & 10.7$^\circ$ & $-$6.7 & $-$2.8 & $-$3.2 \\
PBE, this study     & 10.1$^\circ$ & $-$7.2 & $-$3.1 & $-$3.4 \\[8pt]
LEED, Ref.~\onlinecite{duke2}
                    & 12$^\circ\pm$5$^\circ$ & $-$3$\pm$6 \\[8pt]
LDA+pw, Ref.~\onlinecite{filippetti}
                    & 11.7$^\circ$ & $-$6.0 &  &  \\
LDA+Gauss, Ref.~\onlinecite{schroer1}
                    &  3.6$^\circ$ & $-$7.9 & $-$5.2 & $-$2.7 \\
HF, Ref.~\onlinecite{jaffe1}
                    &  2.3$^\circ$ & $-$7.2 & $-$3.6 & $-$3.4 \\
B3LYP, Ref.~\onlinecite{wander1} + \onlinecite{comment3}
                    &  5.2$^\circ$ & $-$4.9 & $-$2.9 & $-$0.5 \\
\end{tabular}
\end{minipage}
\end{center}
\caption{\label{tab:1010}
Tilt angle $\omega$ of the surface dimer (see Fig.~\protect\ref{fig:1010})
and relative bond length contraction $C_{\rm B}$ (in \% of the corresponding
bulk value) of the surface bonds for the nonpolar (10$\bar{1}$0) surface in
comparison with LEED experiment and previous calculations.
$C_{\rm B,\parallel}$ refers to the Zn--O dimer bond parallel to the surface,
$C_{\rm B}$(Zn) to the back bond of zinc to oxygen in the second layer, and
$C_{\rm B}$(O) to the respective back bond of the surface O atom. Bulk values
of the surface and the back bonds are $u\,c$ and
$\big( (1/2\!-\!u)^2+a^2/3c^2\big)^{1/2}c$, respectively.}
\end{table}


\section{Results and discussion}
\label{sec:results}

\subsection{The nonpolar (10$\mathbf\bar{1}$0) and (11$\mathbf\bar{2}$0)
            surfaces}
\label{sec:nonpolar}

The nonpolar wurtzite (10$\bar{1}$0) surface and the closely related
zincblende (110) surface have been studied experimentally and theoretically
for a wide range of III--V and II--VI compound semiconductors. It was found
that all surfaces show the same basic relaxations with the surface cation
moving inwards and the anion staying above, resulting in a tilt of the surface
anion--cation dimers, and the magnitude of the relaxation is determined by a
competition between dehybridizaton and charge transfer
effects.\cite{duke5,duke6,pollmann,filippetti}

At the surface (this applies also to the (11$\bar{2}$0) surface), the
coordination of the surface atoms is reduced from 4--fold to 3--fold,
thereby creating an occupied dangling bond at the anion and an empty
dangling bond at the cation. Two limiting cases may now be distinguished:
In a dominantly covalent bonded compound the cation will rehybridize from
sp$^3$ to sp$^2$ and will move downwards until it lays nearly in the plane
of its three anion neighbors. The anion stays behind (often even an outward
relaxation is observed) tending toward p--like bonds to its neighbors. The
result is a strong tilt of the surface anion--cation dimer (up to 30$^\circ$
are observed) with only a small change of the bondlength. In a dominantly
ionic solid, electrostatics prevails over dehybridization effects. To obtain a
better screening, both, anion and cation, move toward the bulk. The tilt of
the anion--cation dimer will be small but the bond length can be significantly
reduced. Therefore, the relaxation of the surface dimers directly reflects
the covalency or ionicity of the chemical bond in the compound of
consideration.

\begin{figure}[!t]
\noindent
\epsfxsize=246pt
\centerline{\epsffile{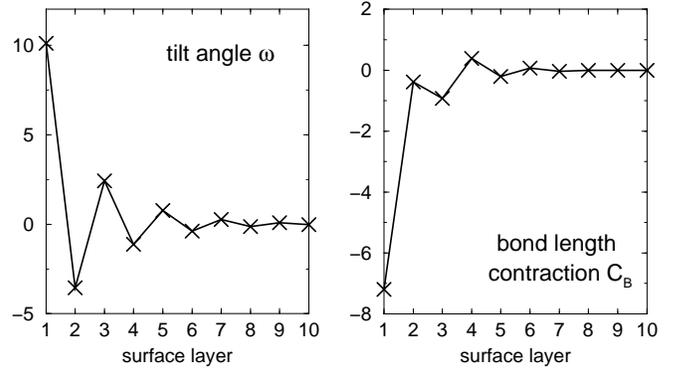}}
\vspace{4pt}
\caption{\label{fig:deeplayer}
Deep-layer relaxations for the nonpolar (10$\bar{1}$0) surface calculated
with a 20 layer slab. Plotted are the tilt angle $\omega$ (in degrees) and the
in-plane bond length contraction $C_{\rm B,\parallel}$ (in \%) of the Zn--O
dimers as a function of the distance from the surface. Only the PBE results
are shown, the results from the LDA calculations are essentially identical.}
\end{figure}

Our results for the relaxation of the (10$\bar{1}$0) surface are given in
Tables~\ref{tab:nonpolar} and \ref{tab:1010}. All lengths are expressed as
fractions of the theoretical lattice parameters given in Table~\ref{tab:bulk}. 
Using these dimensionless relative quantities no significant differences
between the LDA and GGA calculations can be seen. For two structural
parameters the decay of the surface relaxations into the bulk is illustrated
in Fig.~\ref{fig:deeplayer}. Compared to the topmost surface layer, the
tilt angle $\omega$ and the in-plane bond length contraction
$C_{\rm B,\parallel}$ of the Zn--O dimers are already much smaller in the
second and the subsequent layers, but still significant deviations from the
bulk structure can be seen as deep as five or six layers below the surface.

The relatively small angle of $\omega\approx 10^\circ$ for the tilt of the
surface Zn--O dimer together with the Zn--O bond contraction of
$C_{\rm B,\parallel}\approx 7$\,\% confirms that the chemical bond in ZnO is
highly ionic but with significant covalent contributions. A tilt of 10$^\circ$
is at the lower boundary of what has been observed for other III--V and II--VI
compounds.\cite{duke6}  Only the nitride semiconductors show tilt angles that
are similarly small.\cite{filippetti}

The calculated surface relaxations in Table~\ref{tab:nonpolar} and
\ref{tab:1010} agree very well with the DFT-LDA study of
Ref.~\onlinecite{filippetti} and with the results from the LEED
analysis.\cite{duke2}  Relative to the central layer of the slab we
find a downward relaxation of the surface atoms of
$\Delta d_\perp$(Zn)=$-$0.36\,{\AA} and $\Delta d_\perp$(O)=$-$0.04\,{\AA}
with a shift parallel to the surface of $\Delta d_\parallel$(Zn)=0.18\,{\AA}
compared to $\Delta d_\perp$(Zn)=$-$0.45$\pm$0.1\,{\AA},
$\Delta d_\perp$(O)=$-$0.05$\pm$0.1\,{\AA}, and
$\Delta d_\parallel$(Zn)=0.1$\pm$0.2\,{\AA} from the LEED experiment.

Rotation angles of $\omega$=2$^\circ$--5$^\circ$ seem anomalously small in
the context of what is known for other compounds. Even for the very ionic AlN
a tilt angle of $\omega$=7.5$^\circ$ has been reported.\cite{filippetti}
The smaller relaxations obtained in Ref.~\onlinecite{schroer1,jaffe1,wander1}
may be due to not fully converged calculations. Very thin slabs were partly
used or only the first one or two surface layers were relaxed. In
Ref.~\onlinecite{wander1} no relaxation of the Zn ions parallel to the surface
was allowed. Also the convergence of the localized basis sets employed in
these studies and the k--point sampling may have been a problem. As a test
we did a slab calculation where we fixed the positions of the atoms at the
bulk positions and allowed only the first surface layer to relax. The tilt
angle $\omega$ then reduces to roughly half of its value. Also coarsening
the k--point mesh results in changes of 2$^\circ$--3$^\circ$ in $\omega$.
Since we did our calculations with two very different pseudopotential
approaches we can exclude any bias caused by the use of pseudopotentials.

Table~\ref{tab:nonpolar} and \ref{tab:1120} show our results for the
relaxation of the (11$\bar{2}$0) surface. The atomic displacements are of
the same order of magnitude as has been found for the (10$\bar{1}$0) surface.
Again, no significant differences between LDA and GGA calculation can be
seen. The tilt of the surface dimers of 7.5$^\circ$ and the
reduction of the Zn--O dimer bond length of about 6\,\% fits nicely into the
picture of ZnO being at the borderline between ionic and covalent solids.

In a hybrid B3LYP study\cite{wander2} much smaller relaxations for the
(11$\bar{2}$0) surface were reported. However, in this study only three
degrees of freedom per surface layer were relaxed. The authors claimed that
the position of the Zn and O ions are constrained by symmetry. This is not
correct. From the two Zn--O dimers in each surface layer, the atoms of one
dimer can move freely in all three Cartesian directions, leading to six
degrees of freedom per surface layer (see Fig.~\ref{fig:1120}). The position
of the second dimer is then determined by the glide plane symmetry (see also
Ref.~\onlinecite{duke5}).

\begin{table}
\noindent
\begin{center}
\begin{minipage}[c]{246pt}
\def\arraystretch{1.2}
\begin{tabular}{lcccc}
 & $\omega$ & $C_{\rm B,\parallel}$ & $C_{\rm B}$(Zn) & $C_{\rm B}$(O)\\
LDA, this study & 7.6$^\circ$ & $-$5.8 & $-$1.4 & $-$1.7 \\
PBE, this study & 7.4$^\circ$ & $-$6.4 & $-$1.5 & $-$1.8 \\
\end{tabular}
\end{minipage}
\end{center}
\caption{\label{tab:1120}
Tilt angle $\omega$ of the surface dimer (see Fig.~\protect\ref{fig:1120})
and relative bond length contraction $C_{\rm B}$ of the surface bonds for
the nonpolar (11$\bar{2}$0) surface. The same notation as in
Table~\protect\ref{tab:1010} is used.}
\end{table}

\subsection{The polar (0001)--Zn and (000$\mathbf\bar{1}$)--O surfaces}
\label{sec:polar}

\begin{figure}[!t]
\noindent
\epsfxsize=246pt
\centerline{\epsffile{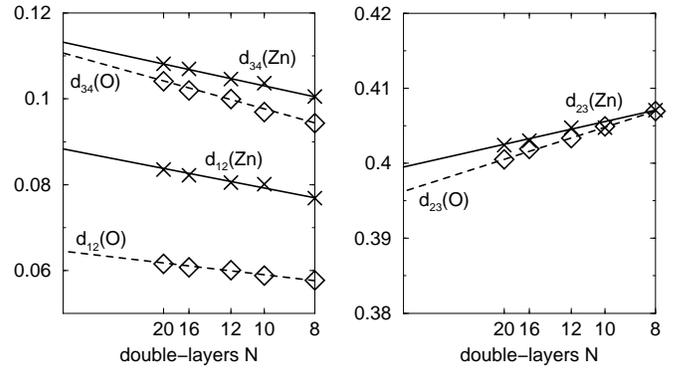}}
\vspace{4pt}
\caption{\label{fig:relax}
First three interlayer distances (see Fig.~\ref{fig:0001}) for the polar
Zn--terminated (0001) and the oxygen terminated (000$\bar{1}$) surface
calculated with different slabs containing $N$ Zn--O double-layers and using
the PBE functional. All distances are given in fractions of the theoretical
bulk lattice constant $c$ (see Table~\ref{tab:bulk}) and are plotted
vs.\ 1/$N$. The extrapolation 1/$N\longrightarrow 0$ gives the surface
relaxations for a vanishing internal electric field.}
\end{figure}

In Figure~\ref{fig:relax} we have plotted the calculated distances between
the topmost surface layers of the polar (0001) and (000$\bar{1}$) surfaces
as a function of the slab thickness $D$. As expected from the thickness
dependence of the residual electric field inside the slab, the $1/D$ plots
reveal a nice linear behavior for the interlayer distances. By extrapolating
$1/D\longrightarrow 0$, all distances may now be obtained in the limit of a
vanishing internal electric field.

The extrapolated results for the relaxations of the polar surfaces are
summarized in Table~\ref{tab:polar} and \ref{tab:0001}. Very good agreement
with the results of previous ab-initio calculations is found. In general,
all double-layers are contracted and the distances between the double-layers
are increased relative to the bulk spacings. For finite slabs, the residual
internal electric field further amplifies this characteristic relaxation
pattern.

The largest relaxation is found for the O--terminated surface where the
outermost double-layer distance is compressed by $\approx$50\,\%. This
agrees reasonably well with the results of the X--ray
experiments\cite{wander3,jedrecy2,jedrecy3} where a contraction of
40\,\%, 54\,\%, and 20\,\% were found. On the other hand, from LEED
analysis\cite{duke4} and LEIS\cite{overbury} measurements it was concluded
that the Zn--O double-layer spacing for the O--terminated surface is close to
its bulk value. The recent finding of W\"oll et al.\cite{woell}  may perhaps
help to solve this contradiction. With helium scattering it was shown that
after commonly used preparation procedures the O--terminated surfaces are
usually hydrogen covered. To test how much hydrogen may influence the surface
relaxations, we repeated a calculation where we adsorbed hydrogen on top of
the O--terminated side of the slab. We find that in this case the outermost
Zn--O double-layer expands again, and the Zn--O separation goes back close to
the bulk distance. A similar result was also reported by Wander and
Harrison.\cite{wander4}

For the Zn--terminated surface there is a clear discrepancy between theory
and experiment. All calculations predict consistently a contraction of the
first Zn--O double-layer of 20--30\,\%, whereas in experiment no
contraction\cite{xpd1} or even an outward relaxation of the topmost Zn--layer
is found.\cite{caiciss,jedrecy2}  This may indicate that the 'metallization'
used in all theoretical studies is not the adequate model to describe the
polar Zn--terminated surface. Recently Dulub and Diebold\cite{diebold}
proposed a new stabilization mechanism for the Zn--terminated surface. With
scanning tunneling microscopy (STM) they found that many small islands with a
height of one double-layer and many pits one double-layer deep are present
on the (0001)--Zn surface. Assuming that the steps edges are O--terminated, an
analysis of the island and pit size distribution yielded a decrease of surface
Zn concentration of roughly 25\,\%. Such a reduction of Zn atoms at the
surface would be enough to accomplish the charge transfer needed to stabilize
the polar surface. It would not be in contradiction with the observed
p(1$\times$1) LEED pattern since a long range correlation between the
different terraces remains. The missing of 25\,\% of the Zn atoms was also
obtained by Jedrecy\cite{jedrecy2} as best fit of their GIXD data.

\begin{table}
\noindent
\begin{center}
\begin{minipage}[c]{246pt}
\def\arraystretch{1.2}
\begin{tabular}{ccccc}
 & \multicolumn{2}{c}{(0001)--Zn surface} &
 \multicolumn{2}{c}{(000$\bar{1}$)--O surface} \\
 & LDA & PBE & LDA & PBE \\ \hline
$d_{12}$ & 0.0952 & 0.0883 & 0.0594 & 0.0645 \\
$d_{23}$ & 0.3947 & 0.3995 & 0.4022 & 0.3962 \\
$d_{34}$ & 0.1172 & 0.1132 & 0.1044 & 0.1107 \\
$d_{45}$ & 0.3811 & 0.3857 & 0.3817 & 0.3784 \\
$d_{56}$ & 0.1187 & 0.1186 & 0.1194 & 0.1251 \\[4pt]
Bulk     & \multicolumn{2}{c}{$(\frac{1}{2}\!-\!u)\,c$ / $u\,c$} &
           \multicolumn{2}{c}{$(\frac{1}{2}\!-\!u)\,c$ / $u\,c$} \\
\end{tabular}
\end{minipage}
\end{center}
\caption{\label{tab:polar}
Summary of the surface relaxations for the polar Zn--terminated (0001) and the
O--terminated (000$\bar{1}$) surface (see Fig.~\protect\ref{fig:0001}). All
distances are given in fractions of the theoretical bulk lattice constant $c$
(see Table~\protect\ref{tab:bulk}).}
\end{table}

\begin{table}
\noindent
\begin{center}
\begin{minipage}[c]{246pt}
\def\arraystretch{1.2}
\begin{tabular}{lcccc}
 & \multicolumn{2}{c}{(0001)--Zn surface} &
   \multicolumn{2}{c}{(000$\bar{1}$)--O surface} \\
 & $\Delta d_{12}$ & $\Delta d_{23}$ &
   $\Delta d_{12}$ & $\Delta d_{23}$ \\ \hline
LDA, this study      & $-$22\,\% & $+$5.1\,\% & $-$51\,\% & $+$4.7\,\% \\
PBE, this study      & $-$27\,\% & $+$5.3\,\% & $-$47\,\% & $+$4.5\,\% \\[8pt]
B3LYP, Ref.~\onlinecite{wander3}
                     & $-$23\,\% & $+$3.5\,\% & $-$41\,\% & $+$3.0\,\% \\
GGA, Ref.~\onlinecite{carlsson}
                     & $-$31\,\% & $+$7.0\,\% & $-$52\,\% & $+$6.5\,\% \\
GGA, Ref.~\onlinecite{noguera}
                     & $-$25\,\% &            & $-$41\,\% & \\
\end{tabular}
\end{minipage}
\end{center}
\caption{\label{tab:0001}
Relaxation of the surface layers of the polar ZnO surfaces in comparison with
previous ab-initio calculations.}
\end{table}

\begin{figure}[!t]
\noindent
\epsfxsize=246pt
\centerline{\epsffile{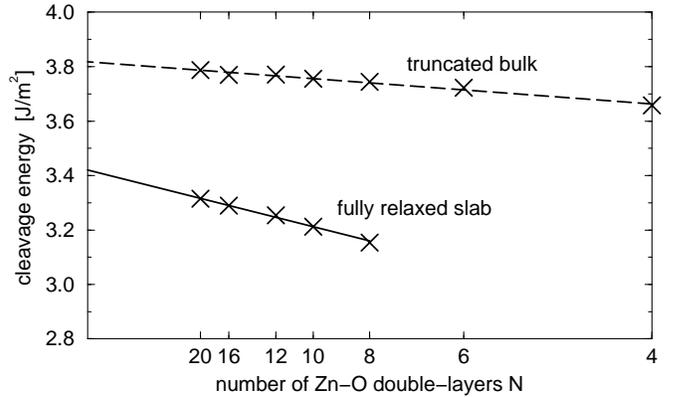}}
\vspace{4pt}
\caption{\label{fig:energy}
Similar plot as Fig.~\ref{fig:relax} for the cleavage energy of the polar
ZnO surfaces. Shown are the results of the PBE calculations.}
\end{figure}

A structure where the surface is stabilized by many small islands and pits
with a Zn deficiency at the step edges is, of course, far away from the model
of a clean, perfectly ordered (0001)--Zn surface used in the theoretical
calculations. Basically all surface Zn--atoms will be next to a step edge,
and therefore very different relaxations may occur. Unfortunately, it is
presently out of the reach of our ab-initio method to do calculations on
slabs representing such an island and pit structure.

For the O--terminated surface, on the other hand, the STM measurements show
a very different picture. Smooth and flat terraces separated mostly by a
two double-layer step are observed. The number of single double-layer steps
was by far not large enough to account for a similar stabilization mechanism
as for the Zn--terminated surface.

\subsection{Surface/cleavage energies}
\label{sec:surfenergy}

For the nonpolar surfaces we can obtain directly the surface energy from our
slab calculations since the slabs are always terminated by the same surface
on both sides. This is not possible for the polar surface since inevitably
both surface terminations are present in a slab calculation. Only the cleavage
energy of the crystal is well defined. To be able to compare the relative
stability of the nonpolar and polar surfaces, we will discuss in the following
only the cleavage energies. The surface energies of the nonpolar surfaces are
just given by half of their cleavage energy.

Like the interlayer distances, the 1/$D$--plot of the cleavage energy for
the polar surfaces in Fig.~\ref{fig:energy} exhibits a simple linear behavior.
As can be seen, the cleavage energy does not change too much with the slab
thickness so that moderate slab sizes would be sufficient to obtain reasonable
converged results.

The extrapolated values for the cleavage energy of the polar surfaces together
with the results for the nonpolar faces and the findings of previous studies
are summarized in Table~\ref{tab:energy}. The nonpolar (10$\bar{1}$0) surface
is the most stable face of ZnO with the lowest cleavage energy. But the energy
of the (11$\bar{2}$0) surface is only slightly higher. The cleavage energy for
the polar surface is roughly a factor of two larger than for the nonpolar
surfaces. This is surprisingly low compared to what has been found in other
systems, for example MgO, where a 'metallization' was also assumed as
stabilization mechanism for the polar surfaces.\cite{pojani}
Therefore, for ZnO the 'metallization' mechanism can well compete with other
stabilization mechanisms like reconstructions or randomly distributed
vacancies and can not be ruled out by energetic considerations alone.

\begin{table}
\noindent
\begin{center}
\begin{minipage}[c]{196pt}
\def\arraystretch{1.2}
\begin{tabular}{lcc}
   &  $E_{\rm cleav}$  &  $E_{\rm relax}$ \\
\hline
(10$\bar{1}$0) surface:\\
\strut\qquad LDA, this study & 2.3 & 0.37 \\
\strut\qquad PBE, this study & 1.6 & 0.37 \\[4pt]
\strut\qquad LDA+pw, Ref.~\onlinecite{filippetti}  & 1.7 & 0.37 \\
\strut\qquad B3LYP, Ref.~\onlinecite{wander1} + \onlinecite{wander3}
                                                   & 2.3 & \\
\strut\qquad HF, Ref.~\onlinecite{jaffe1}          & 2.7 & 0.38 \\
\strut\qquad Shell model, Ref.~\onlinecite{catlow} & 2.0 \\[8pt]
(11$\bar{2}$0) surface:\\
\strut\qquad LDA, this study & 2.5 & 0.29 \\
\strut\qquad PBE, this study & 1.7 & 0.30 \\[8pt]
(0001)/(000$\bar{1}$) surface:\\
\strut\qquad LDA, this study & 4.3 & 0.28 \\
\strut\qquad PBE, this study & 3.4 & 0.28 \\[4pt]
\strut\qquad B3LYP, Ref.~\onlinecite{wander3} & 4.0 & \\
\end{tabular}
\end{minipage}
\end{center}
\caption{\label{tab:energy}
Cleavage energy $E_{\rm cleav}$ (in J/m$^2$) and relaxation energy
$E_{\rm relax}$ (in eV per surface Zn--O dimer) for the different ZnO
surfaces and in comparison with previous calculations.}
\end{table}

The LDA and GGA results in Table~\ref{tab:energy} show the same ordering for
the cleavage energies of the different surfaces but the absolute GGA energies
are roughly 30\,\% lower than the LDA results. This is a well known
improvement of the GGA, where a much better description of the rapidly
decaying charge density into the vacuum region is achieved. The cleavage
energies agree well with previous theoretical results as given in the Table. 
Surprisingly, the results of the hybrid B3LYP studies are much closer to LDA
than to the GGA results. Interestingly, the relaxation energy is roughly the
same for all surfaces when normalized to one Zn--O pair. This means that
despite the partially filled bands at the polar surfaces, the strength of the
relaxation is almost the same as for the isolating nonpolar faces.


\section{Summary and conclusions}
\label{sec:summary}

A first-principles density-functional pseudopotential approach was used to
determine the fully relaxed atomic structures and the surface/cleavage
energies of the nonpolar (10$\bar{1}$0) and (11$\bar{2}$0) surfaces and the
polar Zn--terminated (0001) and the O--terminated (000$\bar{1}$) basal surface
planes of ZnO.

The main results of the presented investigation are an extensive set of
reliable data for the structural parameters and the energetics of the various
ZnO surfaces within the LDA and the PBE approximation, which we consider to be
a reference for future studies (see in particular the compilations in
Tables~\ref{tab:nonpolar}, \ref{tab:polar}, and \ref{tab:energy}).

For the nonpolar surfaces we could resolve the discrepancy between experiment
and several previous ab-initio studies by showing that if calculations are
carefully converged a moderate tilt of the Zn--O surface dimers with a 
relatively strong contraction of the dimer bond length is obtained. Such a
relaxation pattern is typical for rather ionic compounds but with strong
covalent contribution to the chemical bonding. Our results are in line with
LEED analysis and fit very well the systematic trends that are observed for
other more or less ionic II--VI and III--V semiconductors.

The polar surfaces can only be stable if a rearrangement of charges between
the Zn-- and the O--terminated surfaces takes place. In our calculations the
polar surfaces were stabilized by allowing the electrons to move from the
(000$\bar{1}$)--O to the (0001)--Zn surface, thereby quenching the internal
electric field. Nevertheless, even for thick slabs a finite residual electric
field is present inside the slabs, which affects the results for the
structural parameters and the surface energy. To get well converged
results in the limit of a vanishing internal electric, we repeated all
calculations with slabs consisting of different numbers of Zn--O double-layers
and extrapolated the results to the limit of an infinite thick slab.

For both polar surfaces we obtain a strong contraction of the outermost
double-layer spacing. This agrees well with experiments for the O--terminated
surface but not for the Zn termination, indicating that the electron transfer
may be not the adequate model to describe the stabilization mechanism of the
polar Zn--terminated surface. Since this is consistently predicted by all
calculations, it is very likely that other mechanisms, such as defect
formation, hydroxylation and/or the mechanism proposed by Dulub and Diebold
might stabilize the (0001)--Zn surface.

Concerning the surface energies, we find very similar values for the two
nonpolar surfaces with a slightly lower value for the (10$\bar{1}$0) surface. 
The cleavage energy for the polar surfaces is predicted to be roughly a factor
of two larger than for the (10$\bar{1}$0) face.


\section{Acknowledgments}

We wish to thank Volker Staemmler, Karin Fink, and Christof W\"oll
for fruitful discussions. The work was supported by SFB~558 and FCI. 




\begin{references}

\bibitem{laser} D.M. Bagnall, Y.F. Chen, Z. Zhu, T. Yao, S. Koyama,
M.Y. Shen, and T. Goto, Appl.\ Phys.\ Lett.\ {\bf 70}, 2230 (1997).

\bibitem{catalysis} J.B. Hansen, {\it Handbook of Heterogeneous Catalysis},
G. Ertl, H. Kn\"otzinger, J. Weitkamp (Eds.), Wiley--VCH, Weinheim, 1997.

\bibitem{cox} P.A. Cox, {\it Transition Metal Oxides: An Introduction to Their
Electronic Structure and Properties}, Clarendon Press, Oxford, 1992.

\bibitem{duke1} C.B. Duke, A.R. Lubinsky, S.C. Chang, B.W. Lee, and P. Mark,
Phys.\ Rev.\ B {\bf 15}, 4865 (1977).

\bibitem{duke2} C.B. Duke, R.J. Meyer, A. Paton, and P. Mark, Phys.\ Rev.\ B
{\bf 18}, 4225 (1978).

\bibitem{duke3}  C.B. Duke, J. Vac.\ Sci.\ Technol.\ {\bf 14}, 870 (1977).

\bibitem{noguera} C. Noguera, J. Phys.: Condens. Matter {\bf 12}, R367 (2000).

\bibitem{woell} Ch. W\"oll et al., to be published.

\bibitem{comment1} Several publications\cite{jedrecy1,catlow,schroer1}
quote an earlier LEED analysis of Duke et al., Ref.~\onlinecite{duke1},
where a smaller relaxation of the top-layer Zn of $-$0.3\,{\AA} and a
larger displacement of O of $-$0.1\,{\AA} was found, leading to a smaller
tilt of the Zn--O dimers. As is stated in Ref.~\onlinecite{duke2},
Ref.~\onlinecite{duke1} is an analysis of the same experimental data, but
the wrong structural bulk model of Ref.~\onlinecite{wyckoff} was used.
Additionally, several conceptual improvements were made in the reanalysis
Ref.~\onlinecite{duke2}. Under these circumstances, the earlier publication
should be disregarded in favour of the results of Ref.~\onlinecite{duke2}.

\bibitem{comment2} The  LEED experiments are sometimes interpreted in
literature as to conclude that the surface dimer distance is {\em expanded}
compared to the bulk situation.\cite{jedrecy1,jaffe1,wander1,filippetti}
In these cases, the authors either refere to the older LEED analysis of
Ref.~\onlinecite{duke1}, or they neglect the lateral displacement
$\Delta d_\parallel$(Zn) or they misinterpret $\Delta d_\parallel$(Zn)
as a shift in the wrong direction. Indeed, the sign convention for
the lateral displacements is not very clear in Ref.~\onlinecite{duke2}, but
from the absolute atomic positions given in the summary of
Ref.~\onlinecite{duke2} it becomes clear that Zn relaxes {\em toward} the
O ions thereby {\em shortening} the Zn--O distance. 

\bibitem{goepel} W. G\"opel, J. Pollmann, I. Ivanov, and B. Reihl,
Phys.\ Rev.\ B {\bf 26}, 3144 (1982).

\bibitem{jedrecy1} N. Jedrecy, S. Gallini, M. Sauvage-Simkin, and R. Pinchaux,
Surf.\ Sci.\ {\bf 460}, 136 (2000).

\bibitem{parker} T.M. Parker,  N.G. Condon,  R. Lindsay,  F.M. Leibsle, and
G. Thornton, Surf.\ Sci.\ {\bf  415}, L1046 (1998).

\bibitem{wang} Y.R. Wang and C.B. Duke, Surf.\ Sci.\ {\bf 192}, 309 (1987).

\bibitem{ivanov} I. Ivanov and J. Pollmann, Phys.\ Rev.\ B {\bf 24}, 7275
(1981).

\bibitem{catlow} L. Whitmore, A.A. Sokol, and C.R.A. Catlow,
Surf.\ Sci.\ {\bf 498}, 135 (2002).

\bibitem{schroer1} P. Schr\"oer, P. Kr\"uger, and J. Pollmann, Phys.\ Rev.\ B
{\bf 49}, 17092 (1994).

\bibitem{jaffe1} J.E. Jaffe, N.M. Harrison, and A.C. Hess, Phys.\ Rev.\ B
{\bf 49}, 11153 (1994).

\bibitem{wander1} A. Wander and N.M Harrison, Surf.\ Sci.\ {\bf 457}, L342
(2000).

\bibitem{filippetti} A. Filippetti, V. Fiorentini, G. Cappellini, and
A. Bosin, Phys.\ Rev.\ B {\bf 59}, 8026 (1999).

\bibitem{wander2} A. Wander and N.M. Harrison, Surf.\ Sci.\ {\bf 468}, L851
(2000).

\bibitem{exp-bulk1} {\it Numerical Data and Functional Relationships in
Science and Technology}, Landolt--B\"ornstein, New Series Group III,
Vol.~17a amd 22a. edited by K.-H. Hellwege and O. Madelung, Springer, New
York, 1982.

\bibitem{exp-bulk2} S.C. Abrahams and J.L. Bernstein, Acta Cryst.\ {\bf B25},
1233 (1969); T.M. Sabine and S. Hogg, Acta Cryst.\ {\bf B25}, 2254 (1969).

\bibitem{exp-bulk3} C.H. Bates, W.B. White, and R. Roy, Science {\bf 137}, 993
(1962); W. Class, A. Ianucci, and H. Nesor, Morelco Rep.\ {\bf 13}, 87 (1966);
J.C. Jamieson, Phys.\ Earth Planet.\ Inter.\ {\bf 3}, 201 (1970).

\bibitem{tusker} P.W. Tusker, J. Phys. C: Solid State Phys.\ {\bf 12}, 4977
(1979).

\bibitem{wander3} A. Wander, F. Schedin, P. Steadman, A. Norris, R. McGrath,
T.S. Turner, G. Thornton, and N.M. Harrison, Phys.\ Rev.\ Lett.\ {\bf 86},
3811 (2001).

\bibitem{carlsson} J.M. Carlsson, Comp.\ Mat.\ Sci.\ {\bf 22}, 24 (2001).

\bibitem{duke4} C.B. Duke and A.R. Lubinsky, Surf.\ Sci.\ {\bf 50}, 605 (1975).

\bibitem{wyckoff} R.W.G. Wyckoff, {\it Crystal Structures}, Vol.\ I, 2nd ed.,
Wiley, New York, 1963, p.\ 111--112.

\bibitem{xpd1} M. Sambi, G. Granozzi, G.A. Rizzi, M. Casari, and E. Tondello,
Surf.\ Sci.\ {\bf 319}, 149 (1994).

\bibitem{caiciss} H. Maki, N. Ichinose, N. Ohashi, H. Haneda, and J. Tanaka,
Surf.\ Sci.\ {\bf 457}, 377 (2000).

\bibitem{jedrecy2} N. Jedrecy, M. Sauvage-Simkin, and R. Pinchaux,
Appl.\ Surf.\ Sci.\ {\bf 162-163}, 69 (2000).

\bibitem{overbury} S.H. Overbury, P.V. Radulovic, S. Thevuthasan, G.S. Herman,
M.A. Henderson, and C.H.F. Peden, Surf.\ Sci.\ {\bf 410}, 106 (1998).

\bibitem{xpd2} M. Galeotti, A. Atrei, U. Bardi, G. Rovida, M. Torrini,
E. Zanazzi, A. Santucci, and A. Klimov, Chem.\ Phys.\ Lett.\ {\bf 222}, 349
(1994).

\bibitem{jedrecy3} N. Jedrecy, S. Gallini, M. Sauvage-Simkin, and R. Pinchaux,
Phys.\ Rev.\ B {\bf 64}, 085424 (2001).

\bibitem{hks} P. Hohenberg and W. Kohn, Phys.\ Rev.\ {\bf 136}, B864 (1964);
W. Kohn and L.J. Sham, Phys.\ Rev.\ {\bf 140}, A1133 (1965).

\bibitem{ca} D.M. Ceperley and B.J. Alder, Phys.\ Rev.\ Lett.\ {\bf 45},
566 (1980).

\bibitem{sic} J.P. Perdew and A. Zunger, Phys.\ Rev.\ B {\bf 23}, 5048 (1981).

\bibitem{pbe} J.P. Perdew, K. Burke, and M. Ernzerhof, Phys.\ Rev.\ Lett.\ 
{\bf 77}, 3865 (1996); Phys.\ Rev.\ Lett.\ {\bf 78}, 1396 (1997).

\bibitem{van-usp} D. Vanderbilt, Phys.\ Rev.\ B {\bf 41}, 7892 (1990).

\bibitem{ksv} R.D. King-Smith and D. Vanderbilt, Phys.\ Rev.\ B {\bf 49},
5828 (1994).

\bibitem{van-pp} D. Vanderbilt, Phys.\ Rev.\ B {\bf 32}, 8412 (1985).

\bibitem{mb} B. Meyer, C. Els\"asser, and M. F\"ahnle, FORTRAN 90 program for
mixed-basis pseudopotential calculations for crystals, Max--Planck Institut
f\"ur Metallforschung, Stuttgart (unpublished).

\bibitem{kmh} C.-L. Fu and K.M. Ho, Phys.\ Rev.\ B {\bf 28}, 5480 (1983).

\bibitem{girard} R.T. Girard, O. Tjernberg, G. Chiaia, S. S\"oderholm,
U.O. Karlsson, C. Wigren, H. Hyl\'en, and I. Lindau, Surf. Sci. {\bf 373},
409 (1997).

\bibitem{sic-pp} D. Vogel, P. Kr\"uger, and J. Pollmann, Phys.\ Rev.\ B
{\bf 54}, 5495 (1996); Phys.\ Rev.\ B {\bf 52}, R14316 (1995).

\bibitem{diss-vogel} D. Vogel, Dissertation, Universit\"at M\"unster,
Germany, 1998.

\bibitem{numrec} W.H. Press, S.A. Teukolsky, W.T. Vetterling, B.P. Flannery,
{\it Numerical Recipes}, Cambridge University Press, New York 1986.

\bibitem{bengtsson} L. Bengtsson, Phys.\ Rev.\ B {\bf 59}, 12301 (1999).

\bibitem{bm} B. Meyer and D. Vanderbilt, Phys.\ Rev.\ B {\bf 63}, 205426
(2001).

\bibitem{mp} H.J. Monkhorst and J.D. Pack, Phys.\ Rev.\ B {\bf 53}, 5188
(1976).

\bibitem{comment3} The data given in Ref.~\onlinecite{wander1} is not
consistent. In Table~\ref{tab:1010} we cite the bond length values of Table~3,
Ref.~\onlinecite{wander1}. Using the atomic displacements listed in Table~1,
Ref.~\onlinecite{wander1} will lead to different results for the back bond
lengths.

\bibitem{duke5} C.B. Duke and Y.R. Wang, J. Vac.\ Sci.\ Technol.\ A {\bf 7},
2035 (1989).

\bibitem{duke6} C.B. Duke, J. Vac.\ Sci.\ Technol.\ A {\bf 10}, 2032 (1992).

\bibitem{pollmann} J. Pollmann, P. Kr\"uger, M. Rohlfing, M. Sabisch, and
D. Vogel, Appl.\ Surf.\ Sci.\ {\bf 104/105}, 1 (1996).

\bibitem{wander4} A. Wander and N.M. Harrison, J. Chem. Phys. {\bf 115}, 2312
(2001).

\bibitem{diebold} O. Dulub and U. Diebold, to be published.

\bibitem{pojani} A. Pojani, F. Finocchi, J. Goniakowski, and C. Noguera,
Surf.\ Sci.\ {\bf 387}, 354 (1997).

\end{references}
\end{document}